\documentclass[aps,prx,10pt,onecolumn, reprint, superscriptaddress,floatfix, nofootinbib,  showpacs,longbibliography]{revtex4-2}

\usepackage[english]{babel}
\usepackage{graphicx}  % Required for inserting images
\usepackage{float}

\usepackage{amsmath,amssymb,amsthm}
\usepackage{bm}% bold math
\usepackage{dcolumn}% Align table columns on decimal point
\usepackage{braket}
\usepackage{physics}
\usepackage{siunitx}

\usepackage{dsfont}
\usepackage[utf8]{inputenc}  
\usepackage[T1]{fontenc}    

\usepackage[mathlines]{lineno}% Enable numbering of text and display math
\usepackage[colorlinks=true,linkcolor=red,citecolor=blue,urlcolor=blue]{hyperref}
\usepackage{comment}

\usepackage[pdftex,dvipsnames]{xcolor}
\usepackage{tikz}
\usetikzlibrary{quantikz2} % or \usetikzlibrary{quantikz} depending on your installation
\usepackage{ulem}

\tikzset{
  bellbox/.style   = {fill=yellow!20,   draw=yellow!70!black,   rounded corners,
                      inner xsep=2pt, inner ysep=6pt},
  lubluebox/.style = {fill=blue!8,  draw=blue!70!black,  rounded corners,
                      inner xsep=2pt, inner ysep=6pt},
  lugreenbox/.style= {fill=green!10, draw=green!70!black, rounded corners,
                      inner xsep=2pt, inner ysep=6pt},
  nobox/.style     = {draw=none, fill=none},
  ket/.style       = {draw=none, fill=none},
  blabel/.style    = {label position=below, yshift=-0.50cm,
                      align=center, font=\footnotesize, text width=3.2cm},
  mlabel/.style    = {label position=below, yshift=-0.7cm,
                      align=center, font=\footnotesize, text width=2.2cm},
}

\begin{document}
\linespread{1.2}

\title{Experimental generation of asymmetric keys from the parity\\ structure of a two-photon six-qubit entangled state}

\author{Sanchari Chakraborti}
\affiliation{Quantum Optics \& Quantum Information Laboratory, Department of Electronic Systems Engineering, Indian Institute of Science, Bengaluru 560012, India}

\author{Soumya Asokan}
\affiliation{Quantum Optics \& Quantum Information Laboratory, Department of Electronic Systems Engineering, Indian Institute of Science, Bengaluru 560012, India}

\author{Sunita Khoth}
\affiliation{Quantum Optics \& Quantum Information Laboratory, Department of Electronic Systems Engineering, Indian Institute of Science, Bengaluru 560012, India}

\author{Sujai Matta}
\affiliation{Quantum Optics \& Quantum Information Laboratory, Department of Electronic Systems Engineering, Indian Institute of Science, Bengaluru 560012, India}

\author{C. M. Chandrashekar}
\affiliation{Quantum Optics \& Quantum Information Laboratory, Department of Electronic Systems Engineering, Indian Institute of Science, Bengaluru 560012, India}

%***************************************************
\vspace{2cm}
%***************************************************

\begin{abstract}
\noindent
We propose and experimentally demonstrate an asymmetric key-generation scheme in which public and private keys are derived from a six-qubit photonic entangled state. A six-qubit entangled state is generated from two photons, each carrying one polarization and two path qubits. The state is prepared in an equal superposition of eight of the sixty-four six-qubit basis states, in which each photon's three qubits form a codeword of the classical $[3,2,2]$ even-parity code, and are distributed between the two parties, Alice and Bob. This ensures, each triple is a uniformly random codeword and the two codewords are locked together through the correlation. The parity constraint is a local stabilizer, so each party verifies it on its own data every round, and eliminates the bit flip errors arising from imperfections, detector dark counts and multiple occupancy before it enters the raw key. A single publicly announced bit, chosen at random from either of the path qubits, allows each party to reconstruct the other's full triple while leaking no information to third party. We realize the desired six-qubit state using a Sagnac-based type-II SPDC source, reporting a Bell–CHSH parameter $S = 2.798\pm0.008$, by implementing gate operation on polarization and path degree's of freedom. This approach provides a physics-based alternative to public-key cryptosystems built on computational hardness assumptions. An $t-$bit string generated from multiple rounds for encryption will be intrinsically random. Without the private key, one needs at least $O(2^{t})$ runs on an $t-$qubit quantum system to find the matching bit to decrypt.

\end{abstract}

\vspace{4cm}
%***************************************************
%%%%% KEYWORDS, TITLE %%%%%

\maketitle
%***************************************************

%%%%% INTRODUCTION %%%%%
\noindent
\section{\label{sec:Intro} Introduction}
The security of a cryptographic system depends on the unpredictability of its secret keys. In conventional cryptographic schemes, the generation of keys typically relies on pseudo-random number generators, where the security of the generated sequence depends on the underlying algorithm and the secrecy of its initial seed\,\cite{nielsenchuang}.  For asymmetric key cryptography which enables various secure transactions without sharing the secret password,  public and private key pairs are generated using mathematically complex problems which are based  on computational assumptions. \\

Quantum cryptography employs superposition and measurement induced randomness in some protocols and entanglement in some protocols to establish secret keys. The security of these keys does not depend on the adversary's computational power\,\cite{gisin2002,pirandola2020, pironio2010,bennett2014,herrero2017}. Entanglement-based quantum key generation is an important class of quantum cryptographic protocols, where the inherent correlations of the entangled quantum states are used for secure key generation and distribution\,\cite{ekert1991,bennett1992}. Unlike prepare-and-measure schemes, entanglement-based protocols offer distinct advantages, including the ability to detect eavesdropping through violation of Bell's inequalities\,\cite{ekert1991} and resistance to intercept-resend attacks.\\

Photonic systems provide a well established platform for the realization of quantum cryptographic protocols based on entanglement because of their relatively low decoherence, room-temperature operation, and compatibility with fiber-optic and free-space communication channels \, \cite{gisin2002,scarani2009}. The generation of high quality entangled photon pairs through spontaneous parametric down-conversion\,(SPDC)\,\cite{Entg_Source_Kwait}, combined with advances in single-photon detection and state engineering has played an important role in the development of photonic platform for quantum communication and quantum cryptography\,\cite{sergei2019,wang2025}. Scaling beyond bipartite entanglement to multipartite photonic entanglement has further broadened the scope of quantum cryptographic applications, enabling protocols such as quantum secret sharing\,\cite{hillery1999,chen2005} and conference key agreement\,\cite{murta2020,proietti2021}.\\

In addition to the use of polarization degree of freedom of photons to represent qubits, photons can be simultaneously controlled in different degrees of freedom, such as path, time-bin, frequency, and orbital angular momentum. By encoding information across multiple degrees of freedom, multi-qubit entangled states using fewer photons have been demonstrated\,\cite{18Qubit_GHZ_Pan2018,6qubit_CMC}.
Recent experiments have also demonstrated the use of photonic quantum walk involving polarization and path degree of freedom of photons for universal quantum computation and multi-qubit entanglement generation\,\cite{6qubit_CMC}.  Four-qubit configuration using two photons with implementation of quantum gate operations has been used for publicly verifiable quantum random number generation and cryptographic key generation\,\cite{4qubit_CMC}.\\

In this work, we propose a secure key-generation protocol based on a parity-structured six-qubit entangled state, and we present its theoretical framework together with a quantum circuit for preparing such states. We demonstrate the scheme experimentally in a six-qubit photonic system using two photons, each carrying three qubits: one encoded in polarization and two in path degrees of freedom (d.o.f.). The photon pair is generated using a Sagnac-based source of polarization entangled photons. Each photon (party) then encodes its two path qubits locally, using single- and two-qubit gate operations at its own station. This prepares a six-qubit state in an equal superposition of a subset of the 64 computational basis states, constrained by a parity-code correlation. After a $Z$-basis measurement, each party checks the parity of its own three-qubit string and discards the rounds that fail. Errors are thus detected round by round from local data alone, without revealing any key bits. Alice then publicly announces one of her path bit information as the public key, and encrypts the message using the remaining two bits that are retained as her private keys. Alice then sends the encrypted message to Bob over an authenticated classical channel. Bob, using his own triple and the public key, decodes all the bits in Alice's triple from the entanglement based correlation logic and thereby decrypts the message.\\

The novelty of the protocol is three-fold --- first, it offers \textit{a key reconstruction mechanism} that gives each party the other's complete measurement record from a single publicly announced bit, second, \textit{key rate doubling} due to the generation of two private bits per round for encryption, and third, the \textit{built-in self-check capability} that removes the in-station faults round by round before generating the raw key, lowering its error rate. In this protocol, the path qubits are retained within the shielded stations of the two parties and are never exposed to an eavesdropper except for the public key bit, which is independent of the secret bits and therefore, reveals nothing. Any attempt to extract information from the polarization qubits in the channel disturbs the correlation of the entangled pair, reducing the measured $S$-value\,\cite{ekert1991, Security_Pawlowski2010} and correspondingly increasing the error rate\,\cite{bennett1992}. Therefore, the security of the protocol rests on the monogamy of the shared entanglement\,\cite{Monogamy_Coffman2000, Monogamy_Koashi2004}. Experimentally, we certify the polarization entanglement through quantum state tomography (QST)\,\cite{QST_Kwait2001, QST_Altepeter2005}, obtaining a fidelity close to one ($\mathcal{F} = 0.987$) with the target Bell state and by a Bell-CHSH test\,\cite{CHSH1969} yielding $\mathcal{S} = 2.798\pm0.008$ evidencing a strong violation of the classical bound of $2$. The only option for eavesdropper will be to guess the random $t-$bit string, requiring  $O(2^{t})$ runs on an $t$-qubit quantum system to find the matching bit to decrypt.\\

The paper is organized as follows: Section.~\ref{sec:Theory} describes the theoretical design of the protocol and Section.~\ref{sec:Expt} presents its experimental implementation and results. In Section.~\ref{sec:Conclusion} we conclude with remarks.

%****************************************************
%%%%% THEORY %%%%%
\noindent
\section{\label{sec:Theory} Theory: Secure Key Generation using six-qubit System}

A six-qubit system with qubits $q_1, q_2, \dots , q_6$ has Hilbert space $\mathcal{H}^{(6)} = \bigotimes_{k=1}^{6} \mathcal{H}_{k}$ of dimension $2^{6}=64$, where $\mathcal{H}_{k}$ is the two-dimensional Hilbert space associated with qubit $q_{k}$. A measurement of the full register in the computational basis returns one of the $64$ possible bit strings $(x_1~x_2~x_3~x_4~x_5~x_6)$ with $x_{i}\in \{0,1\}$. The six qubits are partitioned into two triples: $(q_{1}, q_{2}, q_{3})$ and $(q_{4}, q_{5}, q_{6})$, held respectively by the two communicating parties, Alice and Bob. The protocol begins by distributing a maximally entangled pair, for example between $q_{1}$ and $q_{4}$ --- any of the four Bell states may be used, the choice being absorbed into a known local unitary operation. The four remaining qubits in the path d.o.f are initialized locally and acted on only by unitaries that never cross the Alice--Bob cut.\\

Within this constraint, a six-qubit state can be engineered as a superposition of a small subset of the $64$ basis states, chosen so that each party's triple forms a codeword of a classical parity code while the two codewords remain locked together through the shared pair. Alice and Bob measure their respective qubits in the $Z$ basis, obtaining the outcome triples $(x_1~x_2~x_3)$ and $(x_4~x_5~x_6)$ respectively. Alice publicly announces the outcome of one of her two local qubits, and encrypts the message $m$ (two message bits $m_1$ and $m_2$) using the remaining two bits. Bob uses the parity-code correlations and his private keys to recover Alice's full triple, and decrypts the message. \\

The choice of local encoding within each triple gives flexibility in protocol design, so the six-qubit register supports a family of protocols differing in redundancy, error detection, and communication cost. Here, we present one member of this family, in which both triples are encoded into the classical even-parity code. 

%%%%%%%%=============================================================
\subsection*{The Protocol}

\begin{figure}[ht]
\centering
\resizebox{0.95\linewidth}{!}{
\begin{quantikz}[row sep=0.25cm, column sep=0.50cm, font=\small]
% --- Qubit-1
    \lstick{$\ket{0}$}
  & \gate{H}
    \gategroup[6,steps=3,style={bellbox},background,
               label style={blabel}]{Bell state preparation \\$(\ket{\Psi^{+}})$}
  & \ctrl{3}
  & \qw
  & \qw
    \gategroup[3,steps=3,style={lubluebox},background]{}
  & \ctrl{2}
  & \qw
  & \meter{}
    \gategroup[6,steps=2,style={nobox},background,
               label style={mlabel}]{$Z$-measurement}
  & \setwiretype{c}\rstick{$x_1$} 
  \\
% --- Qubit-2
    \lstick{$\ket{0}$}
  &  \qw       
  &  \qw
  &  \qw
  &  \gate{H} 
  &  \qw      
  & \ctrl{1}
  & \meter{}
  & \setwiretype{c}\rstick{$x_2$} 
  \\
% --- Qubit-3
    \lstick{$\ket{0}$}
  &  \qw
  &  \qw
  &  \qw
  &  \qw
  &  \targ{} 
  &  \targ{}
  &  \meter{} 
  &  \setwiretype{c}\rstick{$x_3$} 
  \\[0.5cm]
% --- Qubit-4
    \lstick{$\ket{0}$}
  & \qw
  & \targ{}
  & \gate{X}
  & \qw
    \gategroup[3,steps=3,style={lugreenbox},background,
               label style={blabel}]{Local unitary \\operations}
  & \ctrl{2}
  & \qw
  & \meter{} 
  & \setwiretype{c}\rstick{$x_4$} 
  \\
% --- Qubit-5
    \lstick{$\ket{0}$}
  & \qw       
  & \qw
  & \qw
  & \gate{H} 
  & \qw        
  & \ctrl{1}
  & \meter{} 
  & \setwiretype{c}\rstick{$x_5$} 
  \\
% --- Qubit-6
    \lstick{$\ket{0}$}
  & \qw
  & \qw
  & \qw
  & \qw
  & \targ{} 
  & \targ{}
  & \meter{} 
  & \setwiretype{c}\rstick{$x_6$}
\end{quantikz}
}
\caption{The quantum circuit preparing the six-qubit state $\ket{\Psi^{(6)}}$. The \textit{yellow} box represents the entanglement generation unit acting on $q_{1}$ and $q_{4}$; the \textit{blue} and \textit{green} boxes respectively represent the local encoder circuits for $U_{A}$ and $U_{B}$. A $Z$-measurement on all six qubits returns one of the $8$ states $\ket{a,c,a\oplus c, \bar{a}, d, \bar{a} \oplus d}$, with $a, c, d \in \{0,1\}$.}
\label{fig: circuit}
\end{figure}
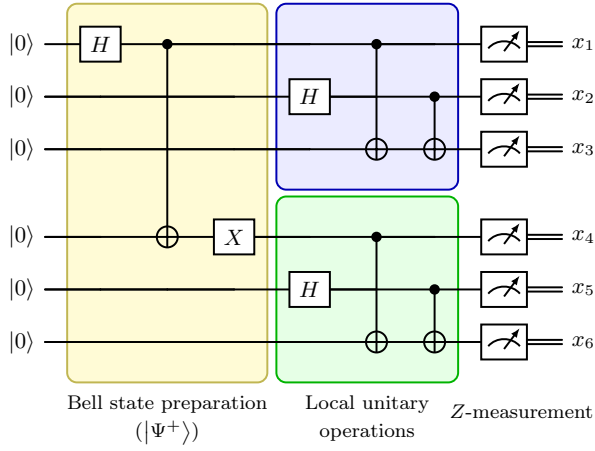

\begin{comment}
    % \lstick{$|0\rangle$} & \gate{H} & \ctrl{3} & \qw      & \qw      & \ctrl{2} & \qw      & \qw  &\meter{} \\
    % \lstick{$|0\rangle$} & \qw      & \qw      & \qw      & \gate{H} & \qw      & \ctrl{1} & \qw  &\meter{} \\
    % \lstick{$|0\rangle$} & \qw      & \qw      & \qw      & \qw      & \targ{}  & \targ{}  & \qw  &\meter{} \\[6pt]
    % \lstick{$|0\rangle$} & \qw      & \targ{}  & \gate{X} &\qw       & \ctrl{2} & \qw      & \qw  &\meter{} \\
    % \lstick{$|0\rangle$} & \qw      & \qw      & \qw      & \gate{H} & \qw      & \ctrl{1} & \qw  &\meter{} \\
    % \lstick{$|0\rangle$} & \qw      & \qw      & \qw      & \qw      & \targ{}  & \targ{}  & \qw  &\meter{}
\end{comment}

%=================================================
\noindent
\paragraph{\textit{Resource and State Engineering:}} 
The protocol is seeded by a maximally entangled pair distributed by a source in the Bell state 
\begin{align}
    \ket{\Psi^{+}}_{14} = \frac{1}{\sqrt{2}} \left( \ket{0}_{1}\ket{1}_{4}+\ket{1}_{1}\ket{0}_{4} \right)~,
    \label{eqn: Psi+}
\end{align}
entangling the qubits $q_1$ and $q_4$, where $q_1$ is retained by Alice and $q_4$ by Bob. The remaining four qubits are initialized locally in $\ket{0}$ within the two secure stations of Alice and Bob and they never enter the channel; only $q_{1}$ and $q_{4}$ traverse the quantum channel and are therefore exposed to eavesdropping.\\

Each party then applies a local encoding unitary to its own qubits: $U_{A}$ by Alice and $U_{B}$ by Bob, where
\begin{subequations}
   \begin{align}
        U_{A} &= CNOT_{2\rightarrow 3}CNOT_{1\rightarrow 3}H_{2}~, \label{eqn: U_A} \\[3pt]
        U_{B} &= CNOT_{5\rightarrow 6}CNOT_{4\rightarrow 6}H_{5}~.
   \end{align}
\label{eqn: unitaries}
\end{subequations}

\noindent
Here $H$ is the Hadamard gate and $CNOT_{i\rightarrow j}$ denotes the controlled-NOT operation with control qubit $q_{i}$ and target qubit $q_{j}$. Applying $U_{A} \otimes U_{B}$ to the initial six-qubit state $\ket{\Psi^{+}}_{14}\ket{0000}_{2356}$ yields
\begin{align}
    \ket{\Psi^{(6)}} = \frac{1}{2\sqrt{2}} \sum_{c,d \in \{0,1\}}  &\Big[ \ket{0}_{1}\ket{c}_{2}\ket{c}_{3}\ket{1}_{4}\ket{d}_{5}\ket{\bar{d}}_{6} 
    \nonumber \\
    &+\ket{1}_{1}\ket{c}_{2}\ket{\bar{c}}_{3}\ket{0}_{4}\ket{d}_{5}\ket{d}_{6} \Big] 
\label{eqn: 6Q_state_Gen}
\end{align}
where $\bar{c} \equiv c\oplus 1$ and $\bar{d} \equiv d\oplus 1$, denoting the complements of $c$ and $d$ respectively. The quantum circuit preparing $\ket{\Psi^{(6)}}$ is shown in Fig.~\ref{fig: circuit}. \\

Equation\,\eqref{eqn: 6Q_state_Gen} is a uniform superposition of exactly $8$ of the $64$ basis states of $\mathcal{H}^{(6)}$,
\begin{align}
    \ket{\Psi^{(6)}} &= \frac{1}{2\sqrt{2}} \Big( \ket{000101}+\ket{000110}+\ket{011101}+\ket{011110} 
    \nonumber \\
    &\quad +\ket{101000}+\ket{101011}+\ket{110000}+\ket{110011} \Big)~.
    \label{eqn: 6Q_state}
\end{align}
The above state can be equivalently written as
\begin{align}
    \ket{\Psi^{(6)}} = \frac{1}{2\sqrt{2}} &\Big[ \left( \ket{000}+\ket{011} \right)_{A} \left( \ket{101}+\ket{110} \right)_{B}
    \nonumber \\
    &+ \left( \ket{101}+\ket{110} \right)_{A} \left( \ket{000}+\ket{011} \right)_{B} \Big]~,
\end{align}
which makes the structure of Alice's and Bob's triples explicit. Each party holds a uniformly random codeword of the classical even-parity code $\mathcal{C} = \{000, 011, 101, 110 \}$, so that $x_{1} \oplus x_{2} \oplus x_{3} = x_{4} \oplus x_{5} \oplus x_{6} = 0$, while the two triples remain anti-correlated through the shared entangled pair, i.e., $x_{1} \oplus x_{4}=1$.  \\

\paragraph{\textit{Correlation structure:}}
The parity constraints established above can be recast in a solved form suited to key reconstruction, giving the following deterministic correlations among the measured bits:
\begin{itemize}
    \item[(i)] $x_{1} = x_{4} \oplus 1$, corresponding to the stabilizer $(-Z_{1}Z_{4})$ of the prepared Bell state $\ket{\Psi^{+}}_{14}~$;
    
    \item[(ii)] $x_{3} = x_{1} \oplus x_{2}$ $~$ and  $~$ $x_{6} = x_{4} \oplus x_{5}$, so that either party's triple is fixed by any two of its bits.
\end{itemize}    
%$x_{1} \oplus x_{2} \oplus x_{3}=0$ corresponding to the stabilizer $Z_{1}Z_{2}Z_{3}$, placing Alice's triple in the classical $[3,2,2]$ even-parity code;
%$x_{4} \oplus x_{5} \oplus x_{6}=0$ corresponding to the stabilizer $Z_{4}Z_{5}Z_{6}$, likewise for Bob;
Together these imply that Bob, holding his own triple $(x_{4},x_{5},x_{6})$, can recover $x_{1}$ directly from (i) and needs one of Alice's two remaining bits to determine the other. \\
  
\paragraph{\textit{Public and private key generation:}}
A $Z$-basis measurement of the full register returns one of the eight possible outcomes $\ket{a,c,a\oplus c, \bar{a}, d, \bar{a} \oplus d}$ supporting $\ket{\Psi^{(6)}}$ given in Eq.\,\eqref{eqn: 6Q_state}, where $a, c, d \in \{0,1\}$ and $\bar{a}=a\oplus 1$. Operationally, Alice and Bob each measure their own three qubits in the $Z$-basis, obtaining the triples $(x_{1}~x_{2}~x_{3})$ and $(x_{4}~x_{5}~x_{6})$, respectively. Alice then randomly selects a position $p \in \{2,3\}$ and publicly announces $(p, x_{p})$ over an authenticated classical channel. The announced bit $x_{p}$ serves as her \textit{public key}, while the remaining two bits $x_{1}$ and $x_{p'}$ are retained as her \textit{private key}, where $p'$ denotes the complementary position in $\{2,3\}$, i.e., $p'\ne p$. \\

The announcement of $x_{p}$ leaks nothing about the secret bit $x_{1}$, since both bits $x_{2}$ and $x_{3}$ are individually uniform and independent of $x_{1}$. Here, $x_{2}$ is a uniformly random bit created by the Hadamard $H_{2}$, and $x_{3} = x_{1} \oplus x_{2}$ is uniformly masked by $x_{2}$, making $x_{3}$ unbiased as well. Conditioned on the announcement $x_{p}$, Eve's uncertainty about the secret bit $x_{1}$ is maximal, i.e., $H(x_{1}|x_{p}) = H(x_{1}) = 1$ bit. Therefore, either bit can be published without compromising $x_{1}$. Randomizing $p$ each round costs nothing and removes any positional pattern from the public announcement record.\\

Bob reconstructs Alice's triple from his own measurement outcomes and the announced bit $x_{p}$, using the correlations described above. The secret bit $x_{1}$ is obtained as $x_{1} = x_{4} \oplus 1$ and the remaining private bit follows as $x_{p'} = x_{p} \oplus x_{1}$. The protocol is completely \textit{symmetric} from both sides: Bob can likewise announce $(q, x_{q})$ with $q\in \{5,6\}$ as the public key, from which Alice reconstructs Bob's private pair as $x_{4} = x_{1} \oplus 1$ and $x_{q'} = x_{q} \oplus x_{4}$, where $q' \ne q$. \\

\paragraph{\textit{Local self check per round:}} 
In each round, both Alice and Bob verify the local parity $x_{1}\oplus x_{2}\oplus x_{3} = 0$ and $x_{4}\oplus x_{5}\oplus x_{6} = 0$, respectively on their own outcomes before any classical communication. Any round failing to satisfy the parity constraint is discarded. This is the built-in error detection feature offered by the classical $[3,2,2]$ code. It flags any odd-weight error i.e., a single or triple bit flip within a station, but cannot detect even-weight errors i.e., double bit flips that map one codeword of $\mathcal{C}$ onto another. Two thirds of the even-weight errors, those that alter $x_{1}$ (or $x_{4}$) resulting in $x_{1}=x_{4}$, contribute to the quantum bit error rate (QBER). A detailed error budget, quantifying the individual contributions, is presented in Appendix.\ref{appndx: Error}. \\

After the $Z$-basis measurement, each party independently evaluates the parity of their respective triples as
\begin{align}
    s_{A} = x_{1}\oplus x_{2}\oplus x_{3},~~ s_{B}=x_{4}\oplus x_{5}\oplus x_{6}
\end{align}
where, $s_{A}, s_{B}\in \{0,1\}$. A round is locally flagged as \textit{pass} if $s=0$ or as \textit{fail} if $s=1$. Over $N$ rounds, Alice and Bob each obtain an $N$-bit flag string: $S_{A} = \big( s_{A}^{(1)}, s_{A}^{(2)}, \dots,  s_{A}^{(N)} \big)$ and $S_{B} = \big(s_{B}^{(1)}, s_{B}^{(2)}, \dots, s_{B}^{(N)}\big)$, computed entirely from local data. The parties exchange these strings with each other over an authenticated classical channel and perform a bit-wise $OR$ operation on the two, $S_{A} \lor S_{B}$ and retain only those rounds for which $s_{A}^{(i)} \lor s_{B}^{(i)} = 0$, i.e., those passing the local parity checks on both sides. Consequently, for all retained rounds the announced information is identical. Importantly, the public announcement of the flag strings $S_{A}$ and $S_{B}$ reveals only the parity of each triple and do not disclose the individual measurement outcomes. Therefore, the parity announcement leaks nothing about the private bits. \\

The local self-check therefore enables round by round error detection, which is a key distinguishing feature of our protocol as compared to the conventional entanglement based QKD protocols. In BBM92, errors are detected only statistically, by sacrificing a random subset of rounds and publicly comparing their key bits. The estimated error rate applies to the key as a whole, while the faulty rounds among the remaining unsampled majority stay individually unidentified until error correction. In our protocol, by contrast, every round carries its own local check: a triple that falls outside the code $\mathcal{C}$ is identified individually from local data alone, without revealing its key bit, and is discarded before it enters the raw key, thereby lowering the error rate. Moreover, since errors on the shared pair --- whether from the source or an eavesdropper --- always produce valid codewords, they never trigger the local flags. A failed check therefore points to an in-station fault rather than a disturbance on the channel. \\

%Faults within the stations are caught round by round as they occur, while channel disturbances register only in the QBER. 

\paragraph{\textit{Encryption and Decryption}}

Each round retained after the self-check provides Alice with a public bit $x_{p}$ and the private bits $x_{1}$ and $x_{p'}=x_{1}\oplus x_{p}$, where $p,p'\in \{2,3\}$ and $p\ne p'$. Bob obtains the triple $(x_{4}, x_{5}, x_{6})$ from which he reconstructs Alice's private bits as $x_{1} = \bar{x}_{4}$ and $x_{p'}= x_{p}\oplus \bar{x_{4}}$.  Alice encrypts a message $m = (m_{1}, m_{2}, \dots, m_{n})$ of length $n$ using her private $n$-bit string and sends the ciphertext together with the public bit string to Bob through an authenticated classical channel. The message bits are encrypted as $E(m_{i}) = m_{i} \oplus x_{1}$ and $E(m_{i+1}) = m_{i+1} \oplus x_{p'}$. Bob decrypts the message using the reconstructed private bits of Alice, as $D(m_{i}) = E(m_{i}) \oplus x_{1}$ and $D(m_{i+1}) = E(m_{i+1}) \oplus x_{p'}$. Therefore, the protocol offers two encryption bits per round, doubling the key rate.\\

% which gives $D(m) = (m \oplus x_{p'}) \oplus (x_{p} \oplus x_{1}) = m \oplus (x_{p'} \oplus x_{p} \oplus x_{1}) = m \oplus 0 = m$.

\paragraph{\textit{Exposure to the eavesdropper:}}
Since encoding unitaries $U_{A}$ and $U_{B}$ are executed inside Alice's and Bob's shielded laboratories, the local qubits ($q_2, q_3$ for Alice and $q_5, q_6$ for Bob) never leave their stations. Only $q_1$ and $q_4$ are exposed to the channel and no information leaves a station other than the classical announcements. Eve therefore has access to the distributed pair in transit and to the public record: the announced bits $(p, x_{p})$, the self-check flags $S_{A}, S_{B}$ and the ciphertext $E(m)$. If instead Alice prepares the entangled pair in her own laboratory and sends the partner qubit to Bob, only $q_{4}$ is exposed to the channel. Placing the source inside a trusted station in this way leaves a single qubit exposed to Eve, requiring only one channel to be monitored for noise and eavesdropping.

%*****************************************************
%%%%% COMPLEXITY %%%%%
%\noindent
%\section{\label{sec:Complexity}Computational Complexity Analysis}

%*****************************************************
%%%%% EXPERIMENT %%%%%
\noindent
\section{\label{sec:Expt} Experiment: Photonic Implementation of the Protocol}

\begin{figure*}[ht]
\centering
\includegraphics[width=\textwidth]{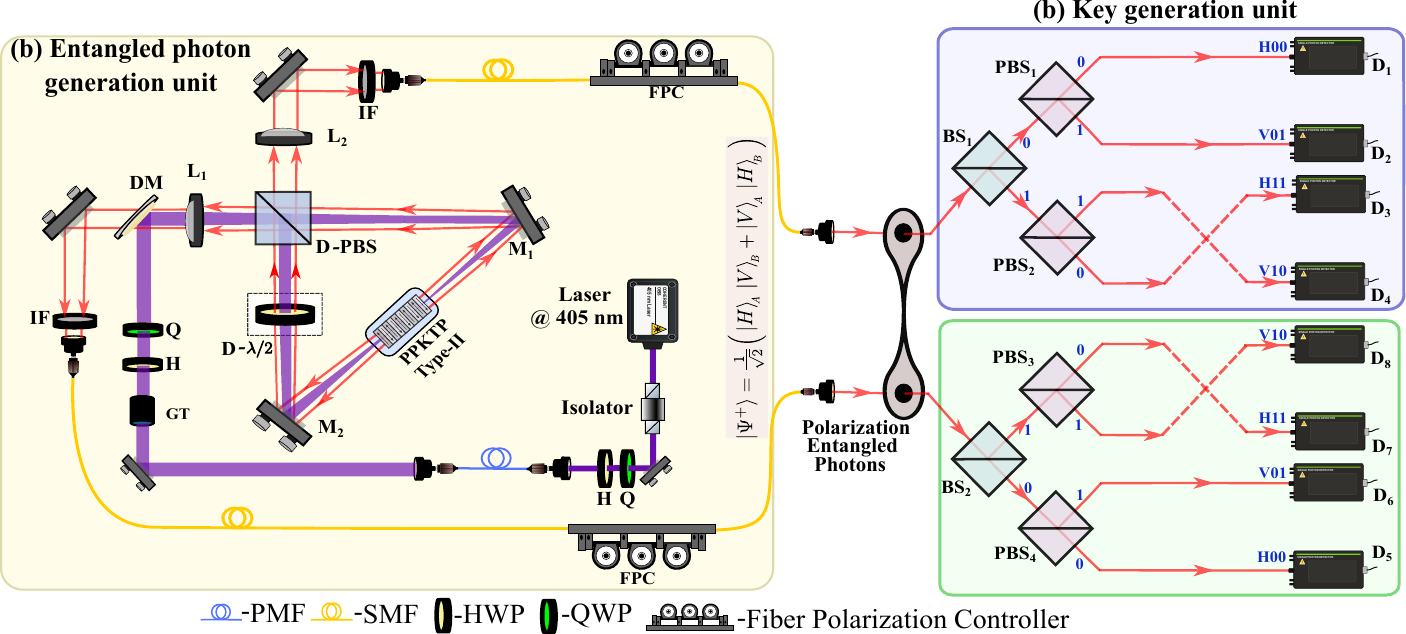}
\caption{Experimental setup. (a) State Preparation: A polarization-entangled state $\ket{\Psi^{+}} = \left( \ket{H}\ket{V}+\ket{V}\ket{H}\right)/\footnotesize{\sqrt{2}}$ is prepared via SPDC process using a type-II PPKTP crystal in a Sagnac interferometer configuration. The photons are coupled into single mode fibers, with fiber polarization controllers compensating for polarization rotation during fiber transmission, thereby maintaining the entangled state. (b) The Key Generation Setup: The polarization entangled photons are sent to the two individual stations of Alice and Bob, followed by path encoding through local unitary operations. The \textit{blue} and \textit{green} boxes show the optical circuits for the local encoders $U_{A}$ and $U_{B}$ with Alice and Bob, respectively. The four output modes of each encoder are directed to four single-photon detectors, so that a detector click identifies one of the four even parity codewords.} 
\label{fig: Expt_schematic}
\end{figure*}

We experimentally implement the protocol on a photonic platform employing two photons, each carrying three qubits distributed over two d.o.f. -- one polarization qubit and two path qubits. Photon A is retained by Alice and photon B by Bob. The joint state space of the two photons is the $64$-dimensional Hilbert space is given by
\begin{align}
    \mathcal{H}^{(6)}_{AB} = \big( \mathcal{H}_{pol}\otimes\mathcal{H}_{p_1}\otimes\mathcal{H}_{p_2} \big)_{A} ~\otimes~ \big( \mathcal{H}_{pol}\otimes\mathcal{H}_{p_1}\otimes\mathcal{H}_{p_2} \big)_{B}
    \label{eqn: HilbertSpace}
\end{align}
where $p_1$ and $p_2$ denote the two path encodings. Each constituent Hilbert space is two-dimensional, with the qubits assigned as $(q_{1}, q_{4}) \rightarrow \mathcal{H}_{pol}$,  $~$ $(q_{2}, q_{5}) \rightarrow \mathcal{H}_{p_1}$ and $(q_{3}, q_{6}) \rightarrow \mathcal{H}_{p_2}$. Here, we adopt the encoding convention $\ket{H} \equiv \ket{0}$ and $\ket{V} \equiv \ket{1}$ for polarization qubit, while the two path qubits together define four spatial modes per photon, labelled as $\{00, 01, 10, 11\}$. Below we present a table-top realization of the circuit shown in Fig.~\ref{fig: circuit}, including the polarization-entangled photon source, the optical encoder circuits and the measurements from which the error and key rates are extracted.

%%%%%%%%=============================================================

\subsection{\label{subsec: state_prep} State Preparation: Source and Local Encoders}

The six-qubit state given in Eq.\,\eqref{eqn: 6Q_state} is prepared in two stages: generation of polarization-entangled photon pairs, followed by local encoding of the two path qubits onto each photon, as shown in Fig.~\ref{fig: Expt_schematic}. A source emitting degenerate pairs of polarization-entangled photons in the Bell state $\ket{\Psi^{+}}$, is realized via SPDC process \cite{SPDC_Couteau} using a type-II PPKTP crystal embedded within a Sagnac interferometer \cite{Entg_Source_Kwait}. At $11$ mW pump power onto the crystal from both sides, we detect a coincidence rate of $342.14 \pm 2.96$ kHz with singles rates of $816.32 \pm 6.79$ kHz and $817.19 \pm 7.04$ kHz, corresponding to a heralding efficiency of $\eta \approx 0.42$ in both the arms. \\

\begin{figure}[h]
\centering
\includegraphics[width=0.999\linewidth, height = 4.3cm]{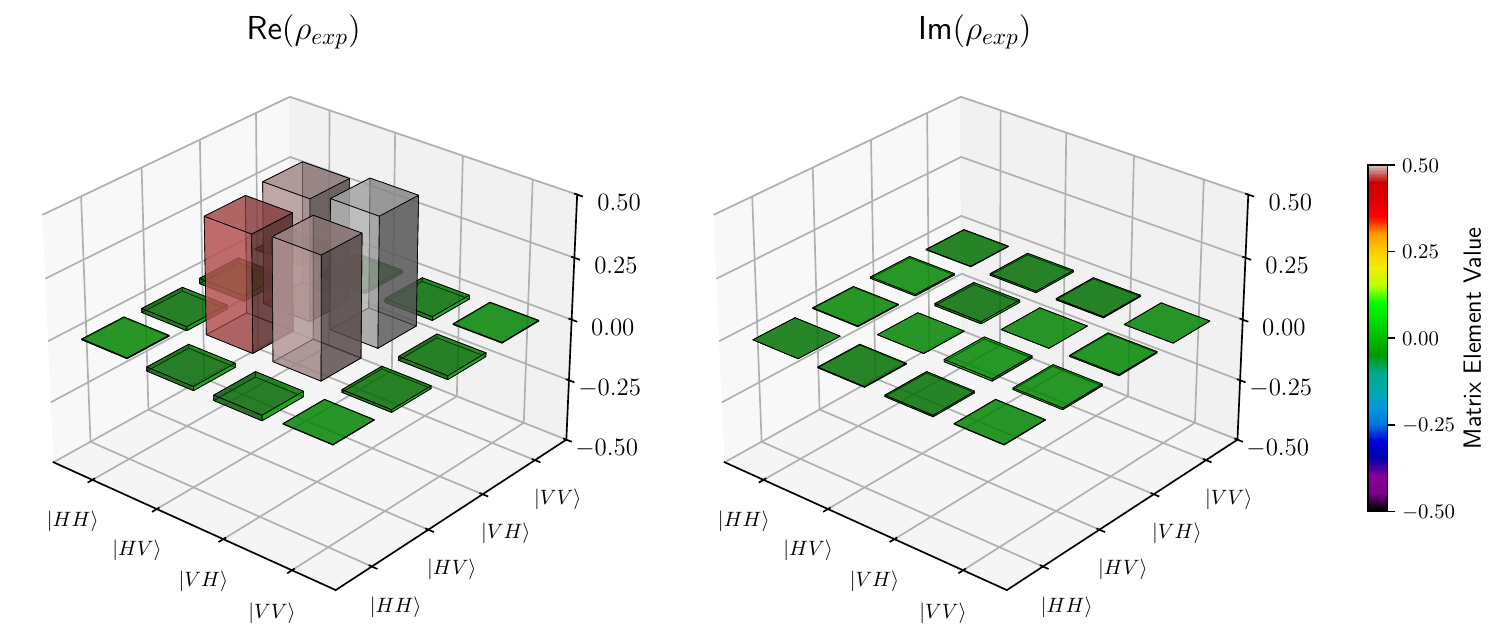}
\vspace{-4mm}
\caption{Real (left) and imaginary (right) parts of the two-photon density matrix $\rho_{_{\mathrm{exp}}}$ reconstructed from polarization-state tomography of the prepared state. } 
\label{fig: tomography}
\end{figure}

The polarization Bell state prepared by the source, $\small{\ket{\Psi^{+}}_{AB} = \frac{1}{\sqrt{2}}\left( \ket{H}_{A}\ket{V}_{B} + \ket{V}_{A}\ket{H}_{B} \right)}$ is characterized by \textit{two-qubit quantum state tomography} \cite{QST_Kwait2001, QST_Banaszek2013}. The reconstructed density matrix $\rho_{\mathrm{exp}}$ is shown in Fig.~\ref{fig: tomography}, with a obtained state fidelity $\mathcal{F} = Tr(\rho_{_{\mathrm{ideal}}}\rho_{_{\mathrm{exp}}}) = \bra{\Psi^{+}}\rho_{_{\mathrm{exp}}}\ket{\Psi^{+}} = 0.987$. The \textit{Bell-CHSH parameter} for the state is measured to be $\mathcal{S} = 2.798 \pm 0.008$, which lies $99.75\sigma$ away from the classical bound $\mathcal{S}_{c}=2$ and within $3.8\sigma$ of the Tsirelson’s bound of $\mathcal{S}_{t} = 2\sqrt{2}$ showing a strong evidence of Bell Violation (details of the source preparation and characterization are provided in the Appendix.~\ref{appndx: Source}). \\

The remaining qubits are encoded locally on each photon through unitary operations implemented with linear-optical elements \cite{LO_ReckZeilinger} acting on the polarization and path degrees of freedom. From the source, the two polarization-entangled photons are directed to separate three-qubit modules, shown as the `blue' and `green' identical boxes in Fig.~\ref{fig: Expt_schematic}, where Alice and Bob apply the local encoding unitaries $U_{A}$ and $U_{B}$. In each module, a balanced beam splitter ($BS$) realizes the Hadamard gate ($H_{2}$, $H_{5}$), placing each photon in an equal superposition of the transmitted (labelled as $0$) and the reflected (labelled as $1$) output modes of the beam splitter, which together define the first path qubit $p_{1}$. The controlled-NOT gates $CNOT_{pol\rightarrow path}$ are realized with polarizing beam splitters ($PBS$) that transmits horizontally polarized light and reflects vertically polarized light, implementing $\ket{H}\ket{0} \xrightarrow{PBS} \ket{H}\ket{0}$ and $\ket{V}\ket{0} \xrightarrow{PBS} \ket{V}\ket{1}$. Therefore, $CNOT_{1\rightarrow 3}$ for photon $A$ and $CNOT_{4\rightarrow 6}$ for photon $B$, writes the polarization bit onto the second path qubit $p_{2}$ so that $x_{3}=x_{1}$ and $x_{6}=x_{4}$. The gate $CNOT_{path\rightarrow path}$ has no single-element realization and is effectively implemented by flipping the target path qubit ($0\leftrightarrow 1$) whenever the control path qubit is $1$, leaving the target qubit untouched when the control is $0$. Therefore, in each module, the spatial modes $10$ and $11$ are interchanged, while $00$ and $01$ remain unaltered. This yields $x_{3} \rightarrow x_{2}\oplus x_{3}$ and $x_{6} \rightarrow x_{5}\oplus x_{6}$. The combined action of $CNOT_{pol\rightarrow path}$ and $CNOT_{path\rightarrow path}$ within the encoders therefore gives $x_{3} = x_{2}\oplus x_{1}$ and $x_{6} = x_{5}\oplus x_{4}$. \\

The four output modes of each module are directed to four single photon detectors, eight in total, $D_{1}$ to $D_{4}$ at Alice's station and $D_{5}$ to $D_{8}$ at Bob's. A click in a detector corresponds to one of the four codewords $\mathcal{C} = \{000, 011, 101, 110\}$ produced by the encoder circuit, giving $(x_1~x_2~x_3) \in \mathcal{C}$ for Alice and $(x_4~x_5~x_6) \in \mathcal{C}$ for Bob. A coincidence between $D_{i} \in [1,4]$ and $D_{j} \in [5,8]$ therefore identifies a unique six-bit string. \\

In the ideal case, the overall setup produces an equal superposition of eight basis states of the six-qubit system, $\ket{\Psi^{(6)}} = \frac{1}{2\sqrt{2}}\sum_{k=1}^{8}\ket{S_{k}}$, given in Eq.~\eqref{eqn: 6Q_state}. However, in practice, the limitations and uncertainties associated with an experiment and the non-ideal optical components result in a mixed state $\rho_{_{\mathrm{exp}}}^{(6)}$, whose population is distributed over the eight states $\ket{S_{k}}$ together with a residual weight outside the codespace,
\begin{align}
    \rho_{_{\mathrm{exp}}}^{(6)} = (1-\epsilon) \rho_{_\mathrm{code}}^{(6)} + \epsilon \rho_{_\mathrm{{err}}}^{(6)}~,
    \label{eqn: 6Q_rho}
\end{align}
where $\rho_{_{\mathrm{code}}}^{(6)}$ is supported on the eight-dimensional codespace $span\{ \ket{S_{k}}\}_{k=1}^{8}$ with populations $p_{k}$ and $\rho_{_{\mathrm{err}}}^{(6)}$ on its orthogonal complement in $\mathcal{H}^{(6)}$. The weight $\epsilon$ is distributed over the remaining $56$ basis states: $8$ error states $span\{\ket{T_{l}}\}_{l=1}^{8}$, in which both triples are valid codewords but $x_{1}=x_{4}$ and $48$ non-codeword outcomes containing at least one odd-parity triple. In our setup, only the error strings $T_{k}$ are registered by the detectors.

%%%%%%%%=============================================================

\subsection{\label{subsec:Expt_ED} Public and Private Key Generation: Methods}
%A $Z$-basis measurement of the full register is inferred from the coincidence detection between the two modules.

%\paragraph{\textit{Data Acquisition:}}
The $Z$-basis measurement is performed locally at each station, and the joint outcome is obtained through coincidence detection between Alice's and Bob's modules. Since each detector registers a single codeword of $\mathcal{C}=\{000,011,101,110\}$, the $4\times4$ detector pairs from both sides produce a total of $16$ outcomes. Eight of these form the \textit{signal} set $S = \{\ket{S_{k}}\}$ i.e., the desired outcomes with $x_{1}\oplus x_{4} = 1$, while the other eight outcomes form the \textit{error} set $T=\{\ket{T_{k}}\}$, with $x_{1}\oplus x_{4} = 0$, i.e., $x_{1}=x_{4}$.
\begin{align}
     S = \{ &\ket{000101}, ~\ket{000110}, ~\ket{011101}, ~\ket{011110},  
    \nonumber \\
    &\ket{101000}, ~\ket{101011}, ~\ket{110000}, ~\ket{110011} \}. \\[6pt]
    T = \{ &\ket{000000}, ~\ket{000011}, ~\ket{011000}, ~\ket{011011},  
    \nonumber \\
    ~&\ket{101101}, ~\ket{101110}, ~\ket{110101}, ~\ket{110110} \}.
\end{align}
In both sets, all the triples are valid even-parity codewords, therefore all $16$ outcomes pass the local self-checks of both parties. \\

The error strings $T_{k}$ form the part of $\rho_{_{err}}^{(6)}$ that is registered experimentally and constitute the $Z$-basis measurement errors, the rate of which defines the QBER denoted as $Q_{Z}$. The populations $p_{k}$ are determined from the recorded coincidence counts $C_{S_{k}}$ between the detector pair ($D_{i}$ at Alice's station and $D_{j}$ at Bob's) corresponding to the signal string $S_{k}$, normalized to the total signal coincidences. Therefore, 
\begin{align}
    p_{k} = \frac{C_{S_{k}}}{\sum_{l=1}^{8} C_{S_{l}}}~,
    \label{eqn: population}
\end{align} 
while the $Z$-basis error rate $Q_{Z}$ is determined as the fraction of all registered coincidences that fall on the error strings $T_{k}$, 
\begin{align}
   Q_{Z} = \frac{\sum_{k=1}^{8} C_{T_k}}{\sum_{k=1}^{8} C_{S_k} + \sum_{k=1}^{8} C_{T_{k}}}~.
   \label{eqn: QBER}
\end{align} 
Experimentally, $Q_{Z}$ captures both the in-station bit flip errors arising from experimental imperfections and the errors originating in the source and from the transmission channel, including multi-pair emission. More details on the distribution of 
the $Z$-basis errors are presented in Appendix.~\ref{appndx: Error}.\\

After the $Z$- measurement, Alice and Bob exchange the respective timestamps of their detection events over the authenticated classical channel to identify coincident events, together with the flag strings $S_{A}$ and $S_{B}$ obtained from the local self-check. In our implementation, this step is enforced by the detection itself in which a non-codeword triple never appears as an outcome. Each coincidence event records one six-bit string $x_{1}~x_{2}~x_{3}~x_{4}~x_{5}~x_{6}$, and $N$ coincidence events produce a measurement record of $N$ strings $\small{ X = \big(x^{(n)}\big)_{n=1}^{N}}$. From the columns of $X$ we extract six $N$-bit strings, $X_{r} = \big( x_{r}^{(1)}, x_{r}^{(2)}, \dots, x_{r}^{(N)} \big)$ for $r=1,2,\dots, 6$, with $X_{1}, X_{2}, X_{3}$ held by Alice and $X_{4}, X_{5}, X_{6}$ by Bob. The raw key formed from these strings therefore contain both \textit{signal} and \textit{error} outcomes. Since the error strings also pass both local self-checks, the error rounds cannot be identified by either party without revealing the key. However, in our analysis, where both records are available, the rounds are sorted into $S$ and $T$ to determine the populations and the QBER directly. \\

In each round $n$, Alice selects a random position $p_{n}\in \{2,3\}$ and announces the positions $P=\{p_{n}\}_{n=1}^{N}$ together with the public key $\small{ X_{P} = \big(x_{p_{n}}^{(n)}\big)_{n=1}^{N}}$, where $x_{2}^{(n)}\in X_{2}$ and $x_{3}^{(n)}\in X_{3}$. Alice retains $X_{1}$ and $\small{ X_{p'} = \big(x_{p'_{n}}^{(n)}\big)_{n=1}^{N}}$ as her private keys, where $p'_{n}$ is the complementary position of $p_{n}$. Bob reconstructs Alices private keys from his own record $X_{4}$ as $X_{1} = \bar{X}_{4}$ and $X_{P}\oplus \bar{X}_{4}$, where $\bar{X}_{4}$ denotes the bit wise complement of $X_{4}$. Rounds belonging to error set $T$ produce mismatches between Alice's private keys and Bob's reconstructions at the rate $Q_{Z}$. Below, we report the experimentally determined raw key rate, QBER and private bit rate.

\subsection{\label{subsec:Expt_results} Results and analysis}

\begin{figure}[t]
\centering
\vspace{-1cm}
\includegraphics[width=0.998\linewidth]{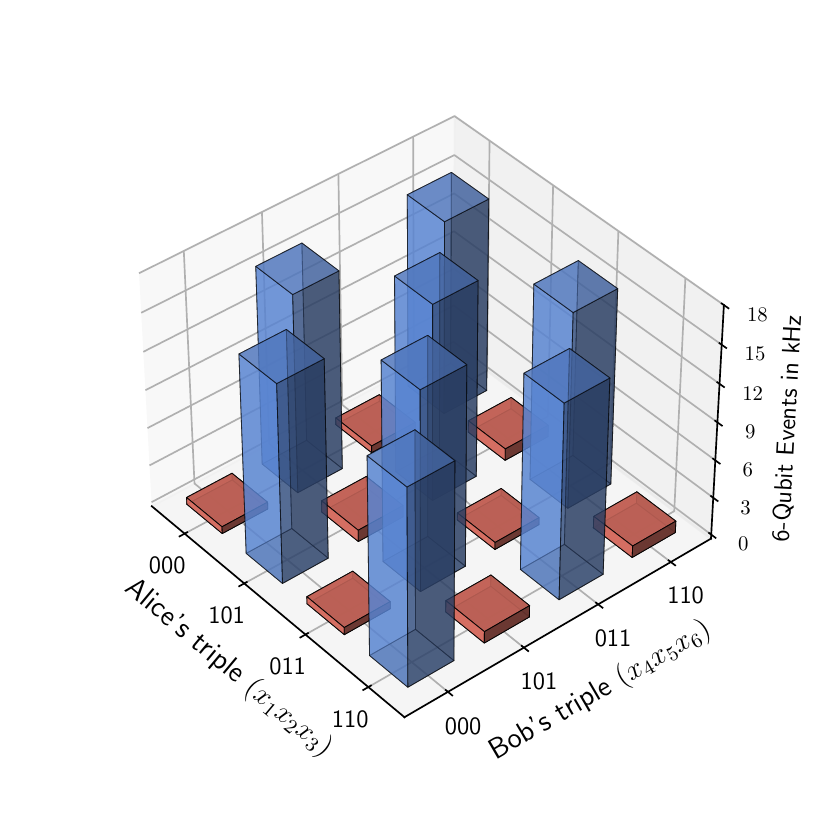}
\vspace{-10mm}
\caption{Measured $Z$-basis coincidence rates for all $16$ six-qubit outcomes registered by the $4\times4$ detector pairs. The $x-y$ axes respectively label the even-parity codewords $\{000,011,101,110\}$ at Alice's and Bob's stations, so that each bar corresponds to one six-bit outcome. \textit{Blue} bars represent the signal set $S$, and the \textit{red} bars represent the error set $T$.} 
\label{fig: Matrix}
\end{figure}

Experimental data are collected by recording coincident events within a $1$ ns coincidence window over an acquisition time of $1$ s and mapping each coincident detection onto its corresponding six-bit string. The total rate of registered coincidences, which gives the raw key rate is obtained as $R_{raw} = 130.925\pm0.114$ kbit/s. Of these, the signal outcomes $S_{k}$ are recorded at $124.885\pm0.112$ kHz, and the error outcomes $T_{k}$ at $6.04\pm0.025$ kHz. Figure~\ref{fig: Matrix} presents the recorded coincidence rates for all $16$ registered outcomes in a $4\times4$ matrix form, showing the eight signal strings (in \textit{blue}) clearly dominating the error strings (in \textit{red}).\\

%\paragraph{results:}  QBER = \frac{error rate}{Total raw keyrate}$\times100$ = $4.61\pm0.018 \%$.

\begin{figure}[h]
\centering
\includegraphics[width=0.998\linewidth]{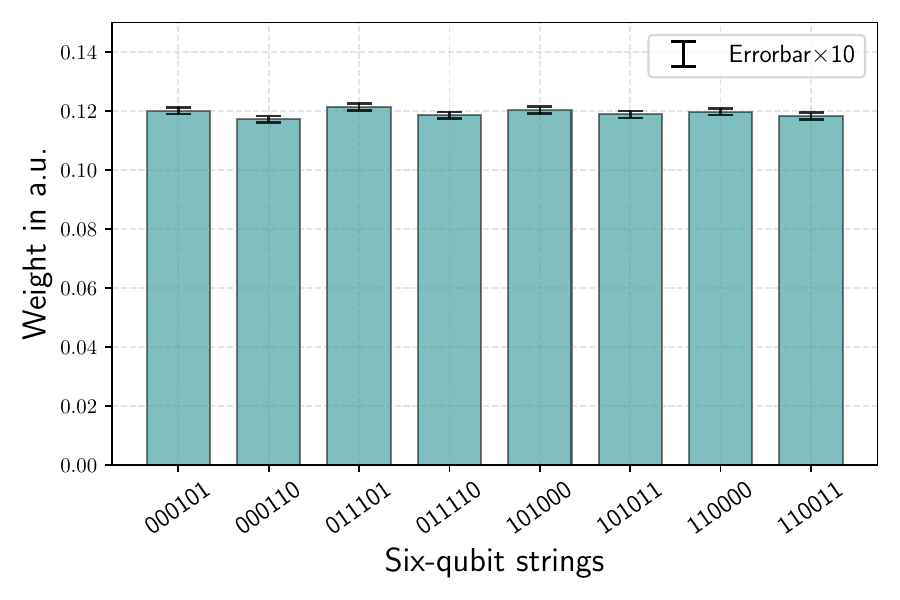}
\vspace{-6mm}
\caption{Population $p_{k}$ of the eight signal strings $S_{k}$ in the raw key, normalized to the total signal counts. The corresponding error bars are multiplied by $10$ for better visibility.} 
\label{fig: 6Q_strings}
\end{figure}

Using Eq.~\eqref{eqn: QBER}, the $Z$-basis error rate is obtained as $Q_{Z} = (4.61\pm0.018)\%$. From the raw coincidences we find the probability of the desired events as $\mathcal{P}_{S} = 0.9538\pm0.0002$ that corresponds to the registered coincidence falling in the signal set $S$, and the probability of the undesired events from the set $T$ as $\mathcal{P}_{T} = 0.0461\pm0.002$, giving a signal-to-noise ratio, $SNR = \mathcal{P}_{S}/\mathcal{P}_{T} = 20.68\pm0.09$. The population $p_{k}$ of the eight signal strings is shown in the bar plot in Fig.~\ref{fig: 6Q_strings}, which is close to the ideal value $1/8 = 0.125$, confirming that the encoders distribute the shared bit uniformly over all the even-parity codewords. \\

The performance of the protocol is characterized by the matrices such as QBER, the key rate and the key symmetry under realistic laboratory conditions. The QBER QBER of $(4.61\pm0.018) \%$ we have recorded for the raw key rate of $130.925\pm0.114$ kbit/s is below the $5\%$ limit, a low error rate for practical key generation. The signal outcomes are recorded at a rate of $124.885\pm0.112$ kHz, from which the private bit rate is obtained as $249.77\pm0.09$ kbit/s. The obtained key symmetries, i.e., the ratios of zeros to ones in the strings $X_{r}$ with $r=1,2,\dots,6$ are $0.998$, $0.993$, $0.990$, $1.002$, $1.020$, $0.993$, respectively, all within $2\%$ of unity, implying uniformly random keys. 
%%%%%%=====================================================================

%*****************************************************
%%%%% CONCLUSION %%%%%
\noindent
\section{\label{sec:Conclusion} Conclusion}
We have proposed an asymmetric key generation protocol built on the parity structure of a six-qubit entangled state. The protocol features round-by-round error detection and offers an alternative to conventional public-key cryptographic schemes. For the experimental demonstration, we configured the six-qubit entangled state using two photons in their polarization and path degrees of freedom. A single polarization-entangled pair, generated from a Sagnac-based source with a fidelity of $0.987$ and a Bell-CHSH parameter of $S = 2.798 \pm 0.008$, was distributed between the distant parties, while the path qubits were encoded locally. This encoding places each party's triple in the classical $[3,2,2]$ even-parity code, with the two codewords locked together through the shared pair. The resulting state is an equal superposition of eight of the $64$ six-qubit basis states, and its deterministic correlations govern the local self-check, the generation of public and private keys, and key reconstruction. The protocol achieves a raw key rate of $130.925 \pm 0.114$~kbit/s at a QBER of $(4.61 \pm 0.02)\%$, with the key bit strings balanced to within $2\%$ of unity. We register the desired signal strings at $124.885 \pm 0.112$~kHz, corresponding to a private-bit rate of $249.77 \pm 0.09$~kbit/s. Beyond generating quantum-secure asymmetric keys from the quantum correlations shared through the polarization degree of freedom and the intrinsic randomness available locally to each party, the demonstration shows that six-qubit states can be engineered deterministically with quantum gate operations in a photonic system using only two photons. \\

The distinguishing features of the protocol arise from its structure. The even-parity constraint provides a built-in self-check: each party can verify the parity of its own triple in every round from local data alone, without revealing the key. Faults within a station are, therefore, removed before they enter the raw key, which lowers its error rate. The self-check also distinguishes faults within the stations from disturbances on the channel, which a single error-rate estimate cannot do. In addition, a single publicly announced path bit, uniformly random and independent of the private bits, allows each party to reconstruct the other's complete triple. The secrecy of the key rests on the monogamy of the shared entanglement, while the path qubits never leave the shielded stations and are therefore never exposed to an eavesdropper. These results establish a physics-based route to asymmetric key generation, in which security derives from quantum resources (intrinsic randomness and multiqubit entanglement) rather than from the presumed computational hardness of mathematical problems that underpins conventional schemes.

%*****************************************************

\vspace{1cm}

\noindent{\bf Acknowledgement}: The authors acknowledge funding support from the National Quantum Mission (NQM), an initiative of the Department of Science and Technology, Govt. of India. S.C. acknowledges Anirudh Verma and Kanad Sengupta for helpful conceptual discussions. \\

\noindent{\bf Data Availability}: The experimental data used to report this study will be available from the corresponding author upon reasonable request.

%*****************************************************
%%%%% APPENDIX %%%%%

%\appendix
%\clearpage
\vspace{1cm}

\appendix
\section*{Appendix}
\label{sec: appendix}
\addcontentsline{toc}{section}{Appendices}
\renewcommand{\thesubsection}{\Alph{subsection}}

\subsection{\label{appndx: Source} Polarization Entangled Photon Source: Preparation and Characterization} 

A bright source of polarization-entangled photon pairs is realized via type-II SPDC in a periodically poled potassium titanyl phosphate (PPKTP) crystal, bidirectionally pumped in a Sagnac geometry by a narrow-linewidth $405$~nm continuous-wave laser. The source generates degenerate photon pairs at $\lambda_{s}=\lambda_{i} =808.58$~nm. As shown in Fig:~\ref{fig: Expt_schematic}, the output of a diode laser [Coherent Obis LX SF] operating at $\lambda_{p}=405$~nm is coupled into a polarization maintaining single mode fiber ($PMSMF$), which spatially filters the pump into a near-Gaussian transverse mode while reducing beam pointing fluctuations. The fiber output is passed through a Glan-Thompson ($GT$) polarizer to obtain a high-purity linear polarization of the pump beam. The polarization state of the pump is subsequently controlled using a half-wave plate (HWP) followed by a quarter-wave plate (QWP), appropriately oriented to prepare the pump in the diagonal polarization state, $\ket{D} = \frac{1}{\sqrt{2}} \left( \ket{H} + \ket{V} \right)$. An achromatic lens ($L_{1}$) of focal length of $20$~cm then focuses the pump beam onto a nonlinear PPKTP crystal placed within a common-path Sagnac interferometer.  \\

The interferometer is built using a dual-band polarizing beam splitter ($D$-$PBS$) and two mirrors ($M_1$, $M_2)$. Upon incidence on the $D$-$PBS$, the diagonally polarized pump is split into two paths according to its polarization. The horizontal ($H$) component is transmitted and propagates clockwise ($CW$) within the interferometer, whereas the vertical ($V$) component is reflected and propagates counter-clockwise ($CCW$). A dual-wavelength half-wave plate ($D$-$\lambda/2$), with its fast axis oriented at $45^{\circ}$ with respect to the horizontal, acts as a $\sigma_{x}$ operator that rotates the pump component in the $CCW$ path from $\ket{V}$ to $\ket{H}$ before it enters the crystal. This ensures that the type-II quasi phase-matching condition is satisfied for both directions of propagation. The two counter-propagating $H$-polarized pump beams are then incident on the $1$~cm long PPKTP crystal, which is housed in a temperature-controlled oven mounted on a 1D-translation stage between $M_1$ and $M_2$. \\

The crystal is periodically poled with a grating period of $10~\mu\text{m}$, chosen to ensure efficient degenerate type-II down-conversion (SPDC). Within the crystal, an $H$-polarized pump photon probabilistically down converts into a pair of daughter photons in $H$ and $V$ polarizations. The pair generated by the clockwise pump passes through the $D$-$\lambda/2$ that changes $H \leftrightarrow V$ at the down-converted wavelength, rendering the two counter-propagating down-conversion processes indistinguishable upon recombination at the $D$-$PBS$. Therefore, the dual-wavelength operation of this half-wave plate within the interferometer is essential for the generation of polarization entanglement. The coherent recombination of the two down-conversion amplitudes at the $D$-$PBS$ results in the polarization-entangled state
\begin{equation}
    \ket{\Psi}_{s,i} = \frac{1}{\sqrt{2}} \left(\ket{H}_{s}\ket{V}_{i} +
    e^{i\phi}\ket{V}_{s}\ket{H}_{i}\right),
    \label{eq:source_state}
\end{equation}
where the subscripts `$s$' and `$i$' denote the signal and idler modes, respectively and the relative phase $\phi = 0$ is set by adjusting the QWP in the pump beam path before the interferometer. \\

\vspace{-2mm}
\begin{figure}[h!]
    \centering
    \includegraphics[width=0.98\linewidth]{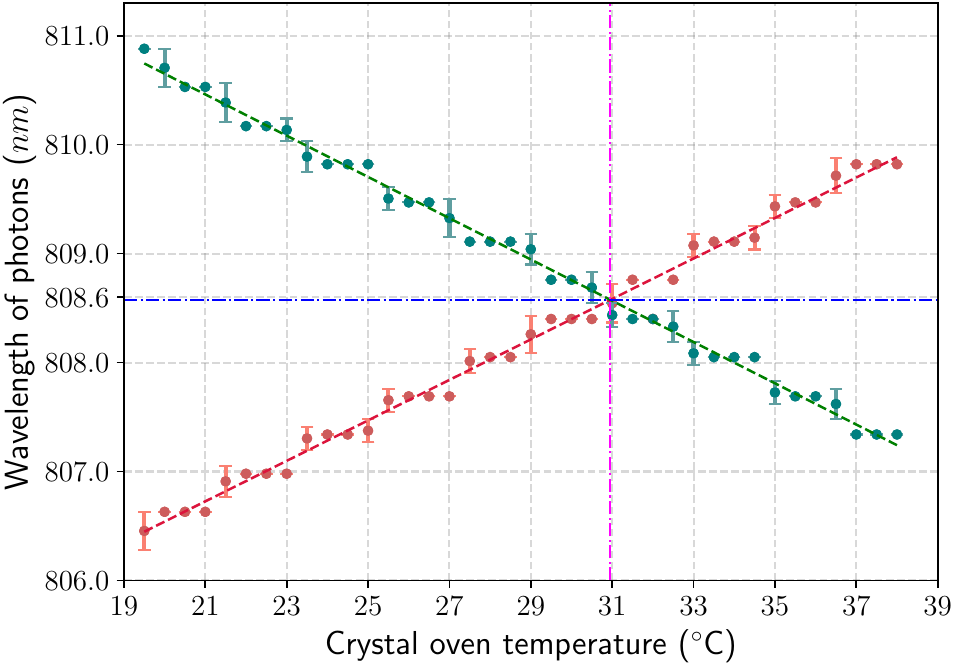}
    \vspace{-2mm}
    \caption{Degeneracy characterization plot for the source: Measured signal and idler wavelengths as a function of the crystal temperature for the type-II PPKTP crystal. The two wavelengths converge at temperature $30.9^{\circ}$C, at which the degeneracy condition $\lambda_{s} = \lambda_{i} = 808.58$~nm is satisfied.}
    \label{fig: degeneracy}
\end{figure}

Degeneracy is achieved by temperature tuning the crystal. The oven temperature is varied from $19.5^{\circ}C$ (room temperature) to $38^{\circ}C$ and the signal and idler spectrum are recorded using a Spectrometer. As shown in Fig.~\ref{fig: degeneracy}, the degeneracy condition $\lambda_{s}=\lambda_{i}$ is met at $30.9^{\circ}$C for this PPKTP crystal and the oven is maintained at this temperature throughout the experiment. \\

The down-converted photons are collimated using lenses $L_{1}$ and $L_{2}$ of focal lengths $20$~cm and collected into single-mode fibers ($SMF$) through fiber couplers [Thorlabs PAF2-7B]. $810\pm10$~nm bandpass interference filters ($IF$) placed before the couplers suppress background light and the residual pump, thereby ensuring spectral purity of the collected photons. Fiber-polarization controllers ($FPC$) are adjusted to compensate for any relative phase introduced due to fiber birefringence. So, at the output end of the two collection fibers, the resultant state becomes %$\ket{\Psi^{+}} = \frac{1}{\sqrt{2}} \left( \ket{H}_{s}\ket{V}_{i} + \ket{V}_{s}\ket{H}_{i} \right)$.
\begin{align}
    \ket{\Psi^{+}} = \frac{1}{\sqrt{2}} \left( \ket{H}_{s}\ket{V}_{i} + \ket{V}_{s}\ket{H}_{i} \right).
    \label{eqn: psi+}
\end{align}
\noindent
which is a maximally entangled state in the polarization degree of freedom of the photons. \\

\paragraph{\textit{State Characterization:}}

Quantum State Tomography (QST) \cite{QST_Kwait2001, QST_Altepeter2005} performed on the prepared state yields a reconstructed density matrix\footnote{The density matrix is reconstructed by maximum-likelihood estimation from the coincidence counts acquired over 36 distinct polarization settings, using Quantum-Tomography Library \cite{KwiatTomo}.}
$\rho_{_{\mathrm{exp}}}$, shown in Fig.~\ref{fig: tomography}. We obtain a state fidelity $\mathcal{F} = \bra{\Psi^{+}}\rho_{_{\mathrm{exp}}}\ket{\Psi^{+}} = 0.987$ with respect to the target Bell state $\ket{\Psi^{+}}$ given in Eq.~\eqref{eqn: psi+}. The concurrence $C = 0.976$ and the negativity $N = 0.974$, both computed from $\rho_{_{\mathrm{exp}}}$, lie close to their maximum value of unity, certifying the high degree of polarization entanglement. \\

\vspace{-2mm}
\begin{figure}[h!]
    \centering
    \includegraphics[width=0.96\linewidth]{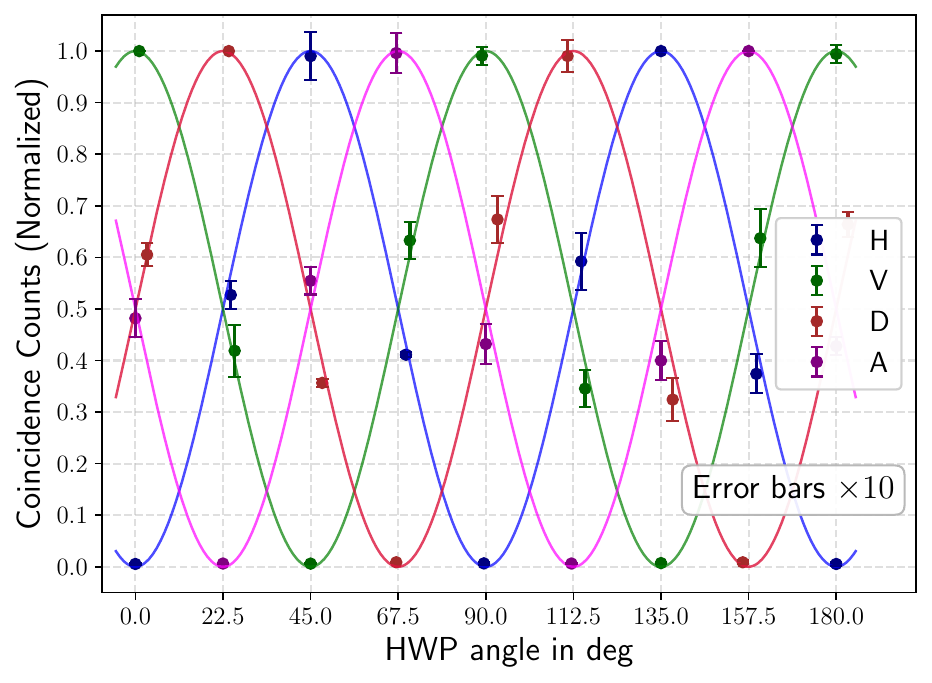}
    \vspace{-2mm}
    \caption{Two-photon coincidence rate as a function of the half-wave plate angle $\theta_{i}$ in idler path, with the signal photon being projected onto $H$, $V$, $D$, $A$ polarizations. The solid curve represents the theoretical expectation, while the dots and bars represent the experimentally obtained mean and $10\times$ standard deviations obtained from photon counting statistics.} 
    \label{fig: Visibility}
\end{figure}

The prepared state is further characterized through two-photon polarization correlations and a Bell-CHSH measurement \cite{CHSH1969}. In the $H/V$ basis, projecting the idler photon onto $\ket{H}$ and $\ket{V}$ while scanning the signal polarization, we obtain visibilities of $98.96\pm0.06\%$ and $98.81\pm0.03\%$, respectively (Fig.~\ref{fig: Visibility}). The corresponding visibilities in the $D/A$ basis, with the idler projected onto $\ket{D}$ and $\ket{A}$, are $98.33\pm0.05\%$ and $98.79\pm0.03\%$, respectively. The near-unit visibility in the superposition basis confirms the coherence of the two-photon polarization superposition, ruling out a classically correlated mixture. \\

\begin{figure}[th!]
    \centering
    \includegraphics[width=0.99\linewidth]{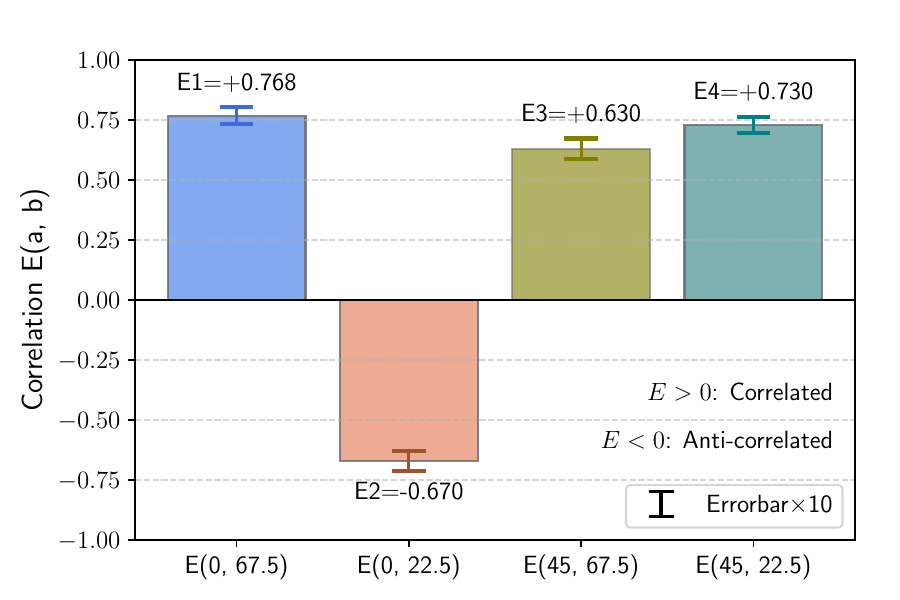}
    \vspace{-4mm}
    \caption{Measured correlations $E(a, b)$ for the prepared state at analyzer settings $a \in \{ 0^\circ, 45^\circ \}$ for photon A and $b \in \{ 22.5^\circ, 67.5^\circ \}$ for photon B, giving a Bell-CHSH parameter $\mathcal{S}= 2.798\pm0.008$. Error bars represent $\pm 1\sigma$ statistical uncertainty, magnified $10\times$ for visibility.} 
    \label{fig: Correlatios}
\end{figure}

The degree of polarization-entanglement is characterized by a Bell-CHSH measurement. The CHSH parameter,
\begin{align}
 \mathcal{S} = \Big| \sum_{x,y \in \{0,1\}} (-1)^{x \cdot y} E(a_{x}, b_{y}) \Big|
\end{align}
is evaluated from the correlators $E(a_{x}, b_{y})$ given as,
\begin{align}
    E(a_{x},b_{y}) = \frac{C_{++}-C_{+-}-C_{-+}+C_{--}} {C_{++}+C_{+-}+C_{-+}+C_{--}} ~,
\end{align}
where $C_{mn} = C(m,n|x,y)$ is the coincidence counts recorded for outcomes $m,n \in \{+, -\}$ at analyzer settings $x$ and $y$.  Here, we choose the analyzer angles $a_0 = 45^{\circ}$, $a_{1} = 0^{\circ}$ for photon A and $b_{0}=67.5^{\circ}$, $b_{1}= 22.5^{\circ}$ for photon B. Figure~\ref{fig: Correlatios} shows the measured correlators, from which we obtain the Bell-CHSH parameter as $\mathcal{S} = 2.798\pm0.008$, exceeding the classical bound $\mathcal{S}\leq 2$ by $99.75\sigma$ statistically, where $\sigma$ is the standard deviation and approaching very close to the Tsirelson bound $\mathcal{S}=2\sqrt{2}$. 

\vspace{-2mm}

% Moreover, to characterize the quality of the prepared state, the visibilities and CHSH-Bell parameter are measured. We obtain the visibilities as $98.962\pm0.057\%$ and $98.807\pm0.029\%$ for the idler projected in $H$ and $V$ polarization, respectively and $98.334\pm0.052\%$ and $98.79\pm0.027\%$ for idler projected in $D$ and $A$ polarization, respectively (Fig.~\ref{fig: Visibility}). The corresponding Bell-CHSH value is obtained to be $\mathcal{S} = 2.798\pm0.008$, demonstrating a strong violation of Bell inequality.\\ 

\paragraph{\textit{Source Statistics:}} 
At a pump power of $11$mW on the crystal, we detect a coincidence rate of $342.14\pm2.96$ kHz with singles rates being $816.32\pm6.79$ kHz and $817.19\pm7.04$ kHz in the signal and idler arms, respectively. This corresponds to a symmetric heralding efficiency $\eta \approx 42\%$ and a detected brightness of $3.11\times10^{4}$ pairs/s/mW. We further measure a heralded second-order intensity correlation function $g^{(2)}_{h}(\tau = 0) = 0.0057\pm0.0001$ for our prepared source, confirming negligible multi-pair emission and the single-photon character of the heralded output.\\

%*****************************************************
%%%%% APPENDIX -2 %%%%%

\subsection{\label{appndx: Error} Quantifying the errors}

We model the errors within each station as independent and identically distributed noise arising from the non-ideal behavior of optical components, detector dark counts, gate infidelity, decoherence, etc. The probabilities of bit flips within a triple are,
\begin{align*}
    \textit{No bit-flip:} ~~~ P_{0} &= (1-p)^{3}\\
    \textit{Single bit-flip:} ~~~ P_{1} &= 3p(1-p)^{2}\\
    \textit{Double bit-flip:} ~~~ P_{2} &= 3p^{2}(1-p)\\
    \textit{Triple bit-flip:} ~~~ P_{3} &= p^{3}   
\end{align*}
where, $p$ is the flip probability per qubit in each station, with the total probability $P_{0}+P_{1}+P_{2}+P_{3}=1$. The odd-weight errors, i.e., single or triple bit flips always change the parity of a codeword and takes a triple out of the code $\mathcal{C} = \{000, 011, 101, 110\}$. Therefore, odd-weight errors are always caught during the parity self-check in individual stations and the corresponding rounds are discarded. On the other hand, the even-weight errors, i.e., the double bit flips map one codeword of $\mathcal{C}$ onto another, leaving the parity unchanged and hence, always remain undetected.  \\
%resulting in bit flips occurring with probability $p$ per qubit. 

The probability that a triple suffers any (at least one) bit-flip within a station is $P = P_{1}+P_{2}+P_{3} = 3p-3p^{2}+p^{3}$. The odd-weight errors, with probability $P_{odd} = P_{1}+P_{3}$, are detected by the self-check and the even-weight errors, with probability $P_{even} = P_{2}$, pass undetected. Therefore, the fraction of errors that escape the local self-check is 
\begin{align}
F_{e}(undetect) = \frac{P_{even}}{P} = \frac{P_{2}}{P_{1}+P_{2}+P_{3}} = \frac{3p(1-p)}{3-3p+p^{2}} ~.
\label{eqn: undet_fraction}
\end{align}
For $p << 1$, undetected errors occur with probability $P_{2} \approx 3p^{2}$ per round and the undetected fraction $F_{e}$ is approximated to $p$. Two thirds of these undetected errors contribute to QBER.  \\

In our experiment, the four detectors in each station are arranged such that a click directly registers one codeword of $\mathcal{C}$, as can be seen from Fig.~\ref{fig: Expt_schematic}. Here, the non-codeword outcomes are rejected by the design of the experimental setup itself and hence, no recorded round fails the local self-check. The only in-station errors that survive are therefore due to the undetected double flips of these even-weight errors, two thirds -- flips on $\{x_{1}, x_{2}\}$ or $\{x_{1}, x_{3}\}$ for Alice and on $\{x_{4}, x_{5}\}$ or $\{x_{4}, x_{6}\}$ for Bob, alter the polarization bit resulting in $x_{1} \oplus x_{4} \ne 1$ and contributes to QBER. To leading order, the $Z$-basis error rate is thus,
\begin{align}
    Q_{Z} \approx \varepsilon_{pair} + 4p^{2}(1-p)~,
    \label{eqn: Z-error}
\end{align}
where, $\varepsilon_{pair}$ is the error rate of the shared entangled pair, arising from the source or in the transition channel before encoding. The second term in $Q_{Z}$ collects the in-station errors that pass the self-check at both stations. The remaining double flip class, i.e., flips on the path qubits $\{x_{2}, x_{3}\}$ for Alice or $\{x_{5}, x_{6}\}$ for Bob, leaves the polarization bit intact and preserves the parity relation $x_{2}\oplus x_{3}=x_{1}$ (or $x_{5}\oplus x_{6}=x_{4}$). Such errors are statistically invisible and affect neither the key nor its reconstruction at the other station. \\

Experimentally, both pair errors ($\varepsilon_{pair}$) and in-station errors produce the same signature, a coincidence with $x_{1} = x_{4}$. So, $Q_{Z}$ is measured directly as the fraction of such coincidences as shown in Eq.\ref{eqn: QBER} and $\varepsilon_{pair}$ is estimated independently from the source characterization. For the prepared entangled state $\ket{\Psi^{+}}$ the expected coincidences are $C_{HV}$ and $C_{VH}$ and therefore, $\varepsilon_{pair}$ can be calculated as,
\begin{align}
    \varepsilon_{pair} &= \frac{C_{HH}+C_{VV}}{C_{HH}+C_{HV}+C_{VH}+C_{VV}} = \frac{1-V_{H/V}}{2} ~. 
\end{align}
Here, $V_{H/V}$ is the two-photon visibility in the $H/V$ basis, defined from the same coincidence counts. Experimentally, we obtained an average visibility of $(98.84\pm0.06)\%$ in $H/V$ basis, which translates to $\varepsilon_{pair} = (0.58 \pm 0.03)\%$. With the measured QBER of $4.56\%$, the undetected in-station errors therefore contribute $Q_{z}-\varepsilon_{pair} = 3.98\%$. Since every recorded click corresponds to a valid even parity codeword, these errors arise from experimental limitations such as the finite extinction ratio of the $PBS$s, polarization dependent reflectivity and transmissivity of the beam splitters, spatial mode cross-talk, detector dark counts, multi-pair emission etc.

%*****************************************************
%%%%% APPENDIX -3 %%%%%

\subsection{\label{appndx: protocol_Choice} Choice of Protocol offered by local encoders}

%(2) \textit{Choice of Protocol}: 
The local gates acting within each triple constitute a design freedom: different encodings produce different logical correlation structures, and hence distinct rules of key reconstruction. The six-qubit register therefore supports a family of protocols rather than a single one, whose members trade redundancy, error-detection capability and classical communication cost against one another, subject to the same entanglement budget of one ebit. The even-parity encoding adopted here spends that freedom in two ways: (i) by imposing a parity constraint on both triples, which each party can verify on its own data every round, and (ii) by keeping one bit of each triple independent of the secret, so that it can be announced publicly to enable reconstruction of the other party's triple. \\

The choice of Bell state used to seed the register is immaterial to the protocol, though it fixes which strings appear in the superposition. For instance, applying the same encoders $U_{A} = CNOT_{2\rightarrow 3}CNOT_{1\rightarrow 3}H_2$ and $U_{B} = CNOT_{5\rightarrow 6} CNOT_{4\rightarrow 6} H_5$ to $\ket{\Phi^{+}}_{14}$ in place of $\ket{\Psi^{+}}_{14}$ preserves the even-parity encoding on each triple. In this case, the resultant state is in a superposition of $\{ 000000, \allowbreak  000011, \allowbreak 011000, \allowbreak 011011, \allowbreak 101101, \allowbreak 101110, \allowbreak 110101, \allowbreak 110110 \}$, with the key bits correlated as $x_1\oplus x_4=0$. Alternatively, a variant protocol maps each triple onto the odd-weight coset $\{ 001, 010, 100, 111 \}$ of the $[3,2,2]$ code, yielding the logical correlations $x_1 \oplus x_2 \oplus x_3 = 1$ and $x_4 \oplus x_5 \oplus x_6 = 1$, with $x_1 \oplus x_4 = 0$ for $\ket{\Phi^{+}}_{14}$ or $1$ for $\ket{\Psi^{+}}_{14}$. In all the above cases, the local parities are deterministic, and the key is reconstructed exactly as before. \\

A genuinely different member of the family is obtained using a repetition encoding, $U_{A} = CNOT_{1\rightarrow 3}~CNOT_{1\rightarrow 2}$ and $U_{B} = CNOT_{4\rightarrow 6}~CNOT_{4\rightarrow 5}~$, which sets $x_{1}=x_{2}=x_{3}$ and $x_{4}=x_{5}=x_{6}$ at each station. A shared entanglement of $\ket{\Phi^{+}}_{14}$ between $q_1$ and $q_4$, therefore results in the six-qubit $\mathrm{GHZ}$ state $\ket{\Psi^{(6)}}_{\mathrm{GHZ}} = \dfrac{1}{\sqrt{2}}\left( \ket{000}_{A}\ket{000}_{B} + \ket{111}_{A}\ket{111}_{B} \right)$, in which each triple is a codeword of the $[3,1,3]$ repetition code. Since the local $Z$-type stabilizers are now $Z_1 Z_2, ~ Z_2 Z_3$ and $Z_4 Z_5, ~Z_5 Z_6$, a single bit flip anywhere within a triple is not just detected but located, enabling recovery of the key bit by majority-vote correction rather than discarding the round. The cost is that no bit of the triple remains independent of the secret, which prevents the parties from publicly announcing anything without revealing the key. \\

The contrast is clean: the $[3,2,2]$ encoding buy a publicly announceable bit at the price of detecting errors without correcting them, while the $[3,1,3]$ repetition code buys error correction at the price of having nothing safe to announce -- both members operate within the same entanglement and qubit budget.

%*****************************************************

\normalem
\bibliography{6Q_KeyGen.bib}

\clearpage

\end{document}